\documentclass[twocolumn,prd,nofootinbib,superscriptaddress,eqsecnum,tightenlines,8pt]{revtex4}

\usepackage{fancyhdr}
\usepackage{amsmath,amsfonts,amssymb}

\usepackage{color}
\definecolor{red}{rgb}{1,0,0}
\definecolor{gre}{rgb}{0,0.6,0}
\definecolor{blu}{rgb}{0,0,1}

\usepackage{mathrsfs}

\usepackage{ulem}

\newcommand{\lb}[1]{\label{#1}}

\newcommand{\be}{\begin{equation}}
\newcommand{\ee}{\end{equation}}
\newcommand{\ba}{\begin{eqnarray}}
\newcommand{\ea}{\end{eqnarray}}

\begin{document}

\title{Our Universe from the cosmological constant}

\date{\today}

\author{Aur\'elien Barrau}
\email{Aurelien.Barrau@cern.ch}
\affiliation{
Laboratoire de Physique Subatomique et de Cosmologie, Universit\'e Grenoble-Alpes, CNRS-IN2P3\\
53,avenue des Martyrs, 38026 Grenoble cedex, France\\
}%

\author{Linda Linsefors}
\email{linda.linsefors@lpsc.in2p3.fr}
\affiliation{
Laboratoire de Physique Subatomique et de Cosmologie, Universit\'e  Grenoble-Alpes, CNRS-IN2P3\\
53,avenue des Martyrs, 38026 Grenoble cedex, France\\
}%


\begin{abstract}
The issue of the origin of the Universe and of its contents is addressed in the framework of bouncing cosmologies, as described for example by loop quantum gravity. If the current acceleration is due to a true cosmological constant, this constant is naturally conserved through the bounce and the Universe should also be in a (contracting) de Sitter phase in the remote past. We investigate here the possibility that the de Sitter temperature in the contracting branch fills the Universe with radiation that causes the bounce and the subsequent inflation and reheating. We also consider the possibility that this gives rise to a cyclic model of the Universe and suggest some possible tests.
\end{abstract}

\maketitle


\section{Basic hypothesis of the model}

 
The scenario presented in this article deals with the question of the origin of the Universe and of its contents in models were a bounce replaces the Big Bang, with a specific emphasis on the loop quantum gravity theory. It does not provide a new model of the high-density universe and relies on successful existing scenarios for the UV behavior. But it does suggest an original geometric understanding of the remote past (in the contracting branch), of the far future (in the expanding branch) and of the origin of matter and radiation. When a bounce replaces the original singularity, those questions become fundamental.\\

There are many attempts to account for the current acceleration of the Universe (see,  {\it e.g.}, \cite{brax} for a pedagogical review). In this work, we will follow the most simple and -- in our opinion -- most natural one : a pure cosmological constant. It should be reminded that in itself the cosmological constant has no link with quantum fluctuations of the vacuum as it is part of standard general relativity. The most general low-energy second order action for the gravitational field, invariant under the diffeomorphisms symmetry is
\begin{equation}
S[g]=\frac{1}{16\pi G}\int (R[g]-2\Lambda)\sqrt{g}, 
\end{equation}
which leads to the Einstein equations with a cosmological constant.
Otherwise stated, in a metric theory of gravity in $d$ dimensions, the generic
local Lagrange density which leads to equations of motion containing
at most second order derivatives of the metric is \cite{Lovelock:1971}:
\begin{eqnarray} \label{Lagrangian}
  L &=& \sum_{n=0}^{[d/2]} c_n L_n \\
  &\equiv& \sum_{n=0}^{[d/2]} c_n 2^{-n} \delta^{\alpha_1\cdots\alpha_{2 n}}_{\beta_1\cdots\beta_{2 n}} R^{\ \ \ \ \beta_1\beta_{2}}_{\alpha_1\alpha_2} \ldots R^{\ \ \ \ \ \ \ \ \ \beta_{2 n - 1}\beta_{2 n}}_{\alpha_{2 n - 1}\alpha_{2 n}}  \nonumber,
\end{eqnarray}

\noindent where $\delta^{\alpha_1\cdots\alpha_k}_{\beta_1\cdots\beta_k}$
is the generalized Kronecker delta symbol, $[d/2]$ is $d/2$ rounded up
to the nearest integer and $c_n$ are constants ($L_0\equiv1$). The first term is the cosmological constant,
the second is the Einstein-Hilbert Lagrange density, and next terms correspond to Lovelock gravity. In $d=4$, Einstein gravity with a cosmological constant is therefore the natural general solution. Nothing here is added "by hands" to fit the observations. 
In the spirit of \cite{rovelli-bianchi}, we assume that the observed acceleration is caused by this $\Lambda-$term and that the quantum fluctuations do not gravitate at all. It is highly probable as the "suppression" factor required to account for the acceleration by the vacuum energy would need to be huge and extremely fine-tuned. And the problem of understanding why the true cosmological constant is exactly zero -- which is not a value favored by general relativity -- would remain. Several solutions to avoid the coupling of quantum fluctuations to gravity are already known, for exemple through deriving gravity by maximizing a suitable entropy functional without using the metric tensor as a dynamical variable \cite{pad}, or by considering the trace-free version of Einstein equations which is essentially equivalent to unimodular gravity \cite{ellis}.\\


The second hypothesis of this article is that, as quite generically expected in quantum gravity, a bounce replaces the Big Bang singularity. Bounces are present in different classes of models, 
{\it e.g.} in the Pre-Big Bang approach \cite{pbb} or in the ekpyrotic model \cite{ek}. This may even happen in classical gravity \cite{patrick}. To remain specific and precise, we however focus in the following on the particular case of the bounce induced by loop quantum gravity (LQG) as this provides a simple, well defined and intensively studied example where a fundamental cosmological contant fits the theory and where the bounce conditions are triggered by the density. It should be made clear that our proposal just requires two assumptions : that the cosmological contant is basically conserved at the bounce and that the bounce itself happens when the density reaches a given critical value. LQG is a non-perturbative and background-invariant quantization of gravity \cite{lqg_review} and loop quantum cosmology (LQC) is its symmetry-reduced version \cite{lqc_review}. Using the Chern-Simons theory in d=3 (that is a topological quantum field theory), successful attempts to account for a cosmological constant through a quantum group structure in d=4 were recently presented \cite{dupuis}, based on the fact that the Turaev-Viro spinfoam model is also defined in terms of quantum groups. In the framework of LQC, it was shown that a simple modification of the amplitude describing the dynamics, corresponding to the introduction of the cosmological constant (and related to the SL(2,C)q extension of the theory considered in \cite{fairbairn}), yields the standard classical de Sitter (dS) cosmological solution \cite{bianchi} in the low energy limit.\\


In LQC, the Big Bang singularity is resolved in a very precise manner due to repulsive quantum geometrical effects (see \cite{abhay} and references therein), the theory being inequivalent to the Wheeler-deWitt approach already at the kinematical level. Holonomy corrections lead to modified Friedmann equations reading (in the high density limit, where the cosmological constant and curvature can be neglected) \cite{lqc_review}:
\begin{equation}
H^2=\frac{8\pi G}{3}\rho\left(1-\frac{\rho}{\rho_{\text{c}}}\right),
\end{equation}
where $\rho_{\text{c}}$ is of order of the Planck density. This equation clearly shows that the Big Bang 
singularity is solved and replaced by a Big Bounce: when $\rho=\rho_{\text{c}}$, the Hubble parameter vanishes 
and changes sign. There is a contracting branch before our current expanding branch. In the remote past, the Universe was necessarily dominated by the cosmological constant: it was just in exponential contraction instead of being in exponential expansion as in the remote future. As the Friedmann equation reads, for a flat $\Lambda-$dominated universe,
$
H^2=\frac{\Lambda}{3},
$
there are indeed positive $H$ and negative $H$ solutions, even for a positive cosmological constant.

In the following, we heavily use the properties of those dS contracting and expanding banches of the Universe. 


\section{The de Sitter horizon}

De Sitter space is the maximally symmetric solution of the vacuum Einstein equations with a positive cosmological constant. It is positively curved with a characteristic length $l=\sqrt{3/\Lambda}$. A key feature is that there are now particle horizons. Each observer is surrounded by a horizon of area $A=4\pi l^2$. In static coordinates, the metric reads
\begin{equation}
ds^2 = -\left(1-r^2\right)dt^2 + \left(1-r^2\right)^{-1}dr^2 + r^2 d\Omega_{n-2}^2.
\end{equation}
It was shown that the close connection between event horizons and thermodynamics found in the case of black holes can be extended to dS spaces \cite{hawking}. 
Any observer perceives a temperature $T$ given by
\begin{equation}\lb{T}
T=\frac{H}{2\pi}=\frac{1}{2\pi l}=\frac{1}{2\pi}\sqrt{\frac{\Lambda}{3}}.
\end{equation}

Many articles were devoted to the thermodynamics of dS spaces \cite{ds}. The current situation is not fully clear. It might be argued that due to the dS temperature the mean value of the stress-energy tensor receives a correction that appears as a modified cosmological constant \cite{markus}. But those results are obtained using standard methods of quantum field theory in curved spaces at vanishing temperature.  The stability of the contracting dS phase was studied in the framework of effective theories and some hints were found in favor of an instability \cite{anderson}. However those analysis were 
not based on the $<in|in>$ formalism  \cite{lee}, which is  more relevant in this case and reaches different conclusions \cite{krotov}. For generic consideration on this issue, see \cite{bousso} and references therein.

The status of particles created by the dS horizon is therefore highly debated (see, {\it e.g.}, \cite{part_ds}). There is no clear consensus. In this work, and this can be considered as our third hypothesis, we assume that they behave as standard radiation that can be considered as a source term in the Friedmann equation, continuously fed by the dS horizon. This is a reasonable and conservative hypothesis supported by the deep analogy between the dS temperature  and the Hawking temperature of a black hole. This is in agreement with (if not required by) all studies of the dS thermal properties (see, {\it e.g.}, \cite{danielsson} and references therein). 
Otherwise stated, backreaction has to be taken into account to go beyond QFT on a curved background and this is mandatory in this case. The universe is in the Bunch Davies vacuum in the remote past. So, either the symmetries are taken seriously and the quantum state is maximally symmetric, just like the dS space, which leads to a correction of the cosmological constant, or the matter and radiation contents are taken seriously and the quantum state then looks like a thermal bath. This second option is the path we follow here as it is -- as explained in the previously quoted articles -- more consistent and arguably more conservative.\\

We therefore consider 
that particles in the dS Universe are, in the spirit of \cite{hawking} and as supported by more recent works \cite{tian}, continuously created by the horizon, in thermal equilibrium at the temperature given by Eq. (\ref{T}). Since it is very low, $T=2\times10^{-34}$ eV, we focus on massless particles. 
As for any gas of thermal bosons the energy and number density are
\be\lb{rho0}
\rho=\frac{\pi^2}{30}g T^4=\frac{g \Lambda^2}{4320\pi^2}~,~n=\frac{\zeta(3)}{\pi^2}g T^3,
\ee
where $g$ is the number of species and $\zeta$ is Riemann zeta function. 
The average total number of particles inside the Hubble horizon will be
$
N=\frac{4\pi l^3}{3}n=\frac{\zeta(3)}{6\pi^4}g=0.0021g,
$
where $g$ is equal to 2 or 4 depending on whether gravitons are included or not. In any case 
$N\ll1$, so that most of the time, the space is truly empty.

\section{Origin of the Universe}
This empty dS space is maximally symmetric : all space-time points are the same, even points in the future and in the past. However, depending on the choice of coordinates, the dS space might appear to either contract or expand. In a pure dS space the difference between expansion and contraction is just a coordinate transformation. The amount of spatial curvature is also just a coordinate choice, but with an upper bound $\frac{k}{a^2}\leq\frac{\Lambda}{3}$ (this can be derived from the Friedmann equation which, in this case, is $H^2=\frac{\Lambda}{3}-\frac{k}{a^2}$). When $\frac{k}{a^2}=\frac{\Lambda}{3}$, the universe will appear to bounce in the considered coordinates. But for any coordinate-invariant observable, a dS universe always remains the same for all times.
If, however, standard fields, {\it e.g.} radiation coming from the dS horizon itself, are introduced, the time symmetry will be broken. In the simplest case, we end up with a homogenous universe that evolves in time. We focus on this specific case to get qualitative ideas of what this implies.\\

We suggest a scenario where the radiation from the dS horizon spontaneously breaks the  symmetry of the dS structure. Out of this, one gets a universe with a tiny amount of radiation, an arbitrary curvature, and a Hubble factor with arbitrary sign. This is fully determined by the emitted radiation which depends on a random quantum process. If the Universe happens to be put in an expanding state by this quantum emission, the radiation will be diluted away and one gets back to an empty dS space until the next radiation is emitted. But if the Universe happens to be contracting, the radiation gets blues-shifted and the energy density of the Universe will increase. 

Let us start with the Friedmann equation (without quantum correction as we here deal with the IR behavior)
\be\lb{fried}
H^2=\frac{\Lambda}{3}-\frac{k}{a^2}+\frac{\kappa}{3}\rho.
\ee
By choosing an appropriate integration constant, the equation of motion for the contracting solution reads
\be
-2\sqrt{\frac{\Lambda}{3}}t=
\ln\left(-K+\sqrt{\frac{\Lambda}{\kappa\rho}}+\sqrt{1-2K\sqrt{\frac{\Lambda}{\kappa\rho}}+\frac{\Lambda}{\kappa\rho}}\right),
\ee
where $K$ is a constant of motion defined as
$
K:=\frac{3k}{2a^2\sqrt{\Lambda\kappa\rho}}.
$ There are tree types of solutions:
\ba
K>1  &&\text{bounce at } -2\sqrt{\frac{\Lambda}{3}}t_B=\ln\left(\sqrt{K^2-1}\right),\\
K=1  &&\text{eternal contraction},\\
K<1  &&\text{crunch at } -2\sqrt{\frac{\Lambda}{3}}t_C=\ln(-K+1).
\ea
One of those cases will be randomly selected by the emission of radiation out of the empty dS space.
If the case $K>1$ is selected, a bounce will happen at the energy density
$
\rho_B=\frac{\Lambda}{\kappa}\left(K+\sqrt{K^2-1}\right)^{-2}<\frac{\Lambda}{\kappa}=1.1\times10^{-123}
$
which is way to small to be the "origin" of the expanding Universe that we observe today. In this case, the radiation will just dilute away.
We conclude that if the dS symmetry is broken so that $H\geq0$ or $H<0$ and $K>1$, the radiation will  just be diluted away. In both those cases,  one gets back to an empty dS space and the process can start again. Another emission will select another case. The probability of ending up with exactly $K=1$ is vanishing. 

Therefore, sooner or later, the Universe will end up with $H\leq0$ and $K<1$.  Here, the energy density will grow to arbitrary large values due to the blue-shift. If we now combine this picture with a quantum bounce, {\it e.g.} from LQC, this sets a suitable origin for our expanding universe.

%

\section{Quantum bounce}


Classically, if $K<1$, which will inevitably happen in this scenario, the Universe ends up in a crunch. But if quantum gravity effets are taken into account, the crunch will be replaced by a bounce. Contrary to the kurvature bounce (which happens when $K>1$), the quantum bounce happens at a high enough energy ($\rho_c\approx0.41$) to be a possible beginning for our expanding universe. At this energy density, the curvature will be
\be
\left(\frac{k}{a^2}\right)_{QB}=\sqrt{\frac{\rho_c}{\rho_0}}\ \frac{k}{a_0^2} \approx-3.3\times10^{123} \ \frac{k}{a_0^2}.
\ee
Since $K<1$ for the quantum bounce to happen, $\left(\frac{k}{a^2}\right)_{QB}<\frac{2}{3}\sqrt{\Lambda\kappa\rho_c}\approx3.6\times10^{-61}$. This sould be compared with $\frac{\kappa}{3}\rho_c\approx3.4$. Therefore, any positive curvature can be safely ignored at the quantum bounce. However there is no similar limit for the negative curvature, which, in principe, can be large.
(Interestingly, the LQC bounce has also been studied in the case of negative curvatures \cite{km}).



The time between the initial symmetry breaking of the dS space and the quantum bounce can be estimated to be
\begin{widetext}
\be
(t_{QB}-t_0)\approx\dfrac{1}{2}\sqrt{\dfrac{3}{\Lambda}}\ln\left[\frac{1}{\textbf{}1-K}\left(-K+\sqrt{\frac{\Lambda}{\kappa\rho_0}}+\sqrt{1-2K\sqrt{\frac{\Lambda}{\kappa\rho_0}}+\frac{\Lambda}{\kappa\rho_0}}\right)\right],
\ee
with interesting limits:
\ba
&\left|\dfrac{k}{a_0^2}\right|\ll\Lambda &\quad\Rightarrow\qquad
(t_{QB}-t_0)\approx\frac{1}{2}\sqrt{\frac{3}{\Lambda}}\ln\left(2\sqrt{\frac{\Lambda}{\kappa\rho_0}}\right)=1.3\times10^{12}\text{ years},
\\
&-\dfrac{k}{a_0^2}\gg\Lambda &\quad\Rightarrow\qquad
(t_{QB}-t_0)\approx\left(-\frac{k}{a_0^2}\right)^{-1/2}.
\ea
\end{widetext}
The duration of the contraction phase can be, depending on the curvature, anything between the Planck time and $1.3\times10^{12}$ years.

When the energy density approaches the critical density, quantum gravity effects enter the game and the bounce takes place. For the specific framework of LQC, we refer the reader to \cite{lqc_review} and references therein for details. Basically, the Wheel-DeWitt equation is replaced by a difference equation : $\partial_\varphi^2 \Psi (\nu, \varphi)= \frac{3\pi G}{4\lambda^2} \nu \left[\, (\nu+2\lambda)
\Psi(\nu+4\lambda) - 2\nu \tilde\Psi(\nu) + (\nu -2\lambda) \Psi(\nu-4\lambda) \right]$ where $\Psi$ is the wave function of the Universe and $ \lambda$ is the square root of the minimum area gap.
 Our scenario elegantly matches the usual LQC view (among others) by providing an explicit mechanism for the origin of the contents that triggers the bounce. The Universe does not need anymore to be arbitrary assumed to be filled by some matter whose origin is mysterious. Here, the origin of the contents is only provided by the quantum geometrical properties of the dS space. 
 
It is natural -- or at least possible -- to assume that during the contraction, the contents of the Universe gets dominated by a scalar field (either fundamental or effective) which would automatically lead to inflation \cite{inflation} and to the usual evolution of the Universe.
The new scenario presented here does {\it not}  replace the usual early Universe evolution and does  {\it not}  depend on the details on this mechanism. The bounce is a very generic feature of quantum geometry (see, {\it e.g.}, \cite{ashtekar2} and references therein) and the detailed contents of the Universe at the bounce time, either a massive scalar field or something else, does not play a significant role in this model. It is of course relevant for the generation of perturbations and the details of the UV behavior but not for the new input of this article which does not deal with particle physics at (or near) the Planck scale. 

The question addressed here is the origin of the content of the Universe and its remote past and future behavior. 



\section{Inhomogeneities}

Throughout this article, we have assumed a homogeneous and isotropic universe. This is certainly not true at all times.\\

By construction of the model, the "initial" state is indeed homogenous and isotropic as it results from the total dilution of any matter contents in the previous expanding branch (the model anyway starts from dS which is by definition homogeneous). It is therefore legitimate to use standard Friedmann equations at this starting point.\\

As soon as a photon is emitted by the horizon and begins to fill the Universe, the homogeneity is broken. Both time and space symmetry are broken. Space symmetry is therefore only assumed at this stage of the development of the model to keep the calculations simple. This assumption should be relaxed in future works. However, some qualitative arguments might lead to think that the main conclusions or this work should remain true. The wavelength of the radiation is indeed initially comparable with the Hubble radius and homogeneity can therefore be assumed to be a reasonable approximation. Spatial homogeneity is not strongly violated. As time goes on, the wavelength gets blue-shifted and the photon is more and more localized. But, in the meanwhile, other photons have been emitted and the Universe remains quite homogeneous when averaged on large enough scales. It is easy to show using Stefan's law that the mean time between two emissions from the horizon is, in Planck units, of order $1/T$. This is precisely the time it takes for the scale factor to change by a sizable amount (that is for the radiation to be substancially blue-shifted). The time needed for the scale factor to vary by a factor $x$ is indeed ${\rm ln}(x)/\sqrt{\Lambda}\sim1/T$. The precise estimate of the inhomogeneities is beyond the scope of this article and should be studied in a future work but using Friedmann equations to describe the basics of the background seems at this stage quite reasonable.

The issue of anisotropies is less severe as, even if they develop and if the shear eventually dominates near the bounce --which is expected--, the main characteristics of the bounce have been already shown to survive.\\

This also raises an interesting feature of this model. In usual LQC, or more generically in bouncing cosmologies, the remote past of the contracting branch (after a possible $\Lambda$-dominated stage) is implicitly assumed to be matter dominated for time-symmetry reasons (the expanding branch is matter-dominated up to the cosmological constant domination phase). This raises a problem. In an expanding matter-dominated universe, it is well known that inhomogeneities grow linearly with the scale factor $a$. Gravity is in competition with the expansion. However, in a contracting universe, both play in the same direction --that is help the growth of inhomogeneities-- and it can easily be shown that perturbations grow (the scale factor is now decreasing) as $1/a^{\frac{3}{2}}$ that is (as expected) faster. Matter inhomogeneities are therefore expected to eventually become very important. On the other hand, in our model, the Universe is only filled by radiation and radiation does not cluster. The initial dS temperature is obviously way too small to lead to the emission of any matter field. As the universe contracts, it will always remain radiation dominated (up to a possible scalar field transition around, {\it e.g.}, the GUT scale). Even after the mean temperature became high enough to create matter particles by scattering, radiation will anyway dominate because of its scale-factor dependance. This circumvents the previously mentioned problem and makes a point for this model. 

\section{Rebirth of the Universe and tests of the model}

Up to now, we have shown how the remote past Universe should be filled with dS radiation due to the cosmological constant which should cause the bounce and the standard evolution. The question of the far future and fundamental origin of the remote past state must now be addressed.
In the future, huge patches of our Universe, with radii larger than the Hubble scale, will be completely empty. They will be pure dS spaces. If the model suggested in this work is correct, these empty spaces will undergo the very same process as described previously (as there is no distinction between an expanding and a contracting pure empty dS space). Radiation will be emitted until one photon randomly leads to a contracting foliation. This makes the model cyclic and solves the question of the origin : the contracting branch emerges from a symmetry breaking of the previous expanding branch (which is neither really expanding or contracting when it becomes pure empty dS). 
It should be emphasized that time always exist in this scenario in the sense of a light cone structure. \\


Is it possible to test this scenario? First, it should be pointed out that no new ``theory" is suggested here. We just link together all the consequences of already accepted or assumed models. The two main ingredients of our proposal are the bounce and the cosmological constant. Both can be tested and, in principle, if both are validated the suggest scenario comes somehow automatically. 
As far as the bounce in concerned, different observational footprints can be expected, even beyond LQC (see, {\it e.g.}, \cite{barrau} and references therein). As far as the interpretation of the acceleration of the Universe by a cosmological constant (or not) is concerned, many experiments are devoted to this issue, in particular the LSST telescope and the Euclid satellite. \\

One step further, this specific scenario of filling the Universe with dS radiation (beyond the bounce and cosmological constant ingredients) can be falsified. Let us consider an example. If our suggestion is correct, one does not expect complex structures in the contracting branch because radiation always dominates. In particular, there is no simple way to form stars and subsequent black holes. However, coalescence of black holes in the contracting phase have been shown to be detectable \cite{ed}. If such circles were to be seen in data, this would disproof our proposal. This statement should be readdressed when inhomogeneities are taken into account.\\

A third insight might come from analog systems. The strong mathematical links between the dS radiation and the Hawking radiation of black holes pointed out in \cite{renaud} could lead to indirect tests of the existence of the dS radiation, as seen in  static coordinates.\\

This simple model builds on the specific properties of dS spaces and bouncing cosmologies to suggest an original new scenario which does not require any assumption about the initial matter contents of the Universe. Everything happens because of the cosmological constant and quantum effects. Particle physics enters the game for the details of the dynamics around the bounce, but the main picture just relies on "vacuum" properties. There are no divergences, no origin of time, and no problem of initial values for the  contents of the Universe. The issue of inhomogeneities and the explicit construction of a global spacetime structure will be studied in future works.\\

\acknowledgments
L.L. is supported by the Labex ENIGMASS.

\end{document}